\documentclass[12pt]{article} %WS 10pt
\title{Scattering Relativity in Quantum Mechanics }  
\author{{\it Richard Shurtleff~}\thanks{affiliation and mailing 
address: Department of Sciences, 
Wentworth Institute of Technology, 550 Huntington Avenue, 
Boston, MA, USA, ZIP 02115, telephone number: (617) 989-4338, fax 
number: (617) 989-4591 , e-mail address: shurtleffr@wit.edu}} 
\begin{document} 
          
\maketitle 

%\hline
\begin{abstract} 

By adding generalizations involving translations, the machinery of the quantum theory of free fields leads to the semiclassical equations of motion for a charged massive particle in electromagnetic and gravitational fields.  With the particle field translated along one displacement, particle states are translated along a possibly different displacement. Arbitrary phase results. And particle momentum, a spin (1/2,1/2) quantity, is allowed to change when field and states are translated. It is shown that a path of extreme phase obeys a semiclassical equation for force with derived terms that can describe electromagnetism and gravitation.

\vspace{0.5cm}


\end{abstract}
%\hline
%\pagebreak

%The remark Eqn Check (N) refers to equation (N) in my notes that check the equations, 7/26/11 - 7/29/11. 

\section{Introduction} \label{intro}
Quantum fields differ from quantum states. While fields are sums of the creation and annihilation operators that add or remove states from multiparticle states, fields transform by nonunitary representations (reps) of spacetime symmetries while  states and their operators transform with unitary reps. 

In a way, states and fields differ somewhat like identical experiments in different labs, say one in lab B and one in lab C. States and fields describe the same physical situation but are somewhat isolated from one another. So, like experiments in labs B and C, there is some flexibility in the reference frames needed to describe the experiment. This is the idea behind `scattering relativity.'
   
One can allowing states and fields to be referenced to different frames. Then the Lorentz transformation from one frame to the other relates any boosts or rotations involved in the description of the experiment. For the semiclassical results obtained here, this freedom is not needed, but it is consistent with the underlying idea and will be kept as part of the calculation. 

If the experiment is moved from one bench to another in lab B and the experiment in lab C is likewise displaced, it should not matter to the outcomes of these identical experiments whether or not the displacements are equal. The relative locations of the equipment involved matters, not where the entire experimental setup is placed. Likewise, it should not matter that fields undergo a different displacement than states. We allow fields to be displaced by a quantity $b$ and states and their operators by a possibly different amount $b_{S}.$ 

Arbitrary displacements bring arbitrariness to the phase, the scalar product of momentum and location $p\cdot x.$ The arbitrariness is shown to introduce general relativistic aspects to the motion such as local and general coordinate systems and Christoffel connections.

The analogy with experiments in labs B and C is not pushed too far. One follows the analogy to generalize a standard process of determining fields. Since these are generalizations, conventional results are the default.

Another generalization involves reps of translations for various spins.  In particular momentum is a 4-vector, with spin (1/2,1/2), and can be changed under a translation by such reps when the momentum is linked to a second rank tensor, with spins (0,0),(1,0), (0,1), and (1,1). The linked tensor is arbitrary, an array of free variables to be constrained by the assumptions. It turns out to be related to the electromagnetic field.

The results of the generalized calculation are evaluated by rudimentary methods appropriate to the introduction of new ideas. Paths are deduced semiclassically by following extreme phase.  

Fields describing massive particles are sums of coefficient functions times operators. In Sec. 2, the generalizations are applied to a conventional process that derives expressions for the coefficient functions. We follow Ref. \cite{W} closely. Paths of extreme phase are obtained in Sec. 3 from the expression for phase in Sec. 2. In Sec. 4, parallel translations of the spin (1/2,1/2) particle momentum are developed, making a  momentum at an initial location equivalent to a possibly different momentum at some other location. The parallel translation of momentum is path dependent. In Sec. 5, the arbitrariness in phase accompanying the dual displacements for states and fields is developed and is shown to describe curved spacetime. The path of extreme phase obeys a semiclassical force equation that is shown in Sec. 6 to correspond to the motion of a charged massive particle in a combined electromagnetic and gravitational field.

\pagebreak

\section{Field and State} \label{2}

This section supplies a derivation of some properties of the quantum field of a massive particle species, following Ref. \cite{W} closely. But fields are allowed to be transformed by reps of translations. We do not agree that fields must be translation invariants.\cite{Raymond,W1} 

The quantum field $\psi_{l}(x)$ for a species of particles of mass $m$ and spin $j$ is constructed as a linear combination of annihilation and creation operators, $\psi_{l}(x)$ = $\kappa \psi^{+}_{l}(x)$ + $\mu \psi^{-}_{l}(x).$ One has an annihilation field $\psi^{+}$ and a creation field $\psi^{-}$ given by 
$$\psi^{+}_{l}(x) =  \int d^3 p \enspace u_{l\sigma}(x,{\overrightarrow{p}}) a_{\sigma}({\overrightarrow{p}})  \, ,$$
\begin{equation} \label{psi+}
\psi^{-}_{l}(x) =  \int d^3 p \enspace v_{l\sigma}(x,{\overrightarrow{p}}) a^{\dagger}_{\sigma}({\overrightarrow{p}})  \, , %Eqn Check (B1,2,3)					
\end{equation}
where the repeated index $\sigma$ is summed, $a_{\sigma}({\overrightarrow{p}})$ and $a^{\dagger}_{\sigma}({\overrightarrow{p}})$ are operators that remove or add an eigenstate of momentum $\overrightarrow{p}$ and spin component $\sigma.$ The spatial components $\overrightarrow{p}$ are free while the time component of momentum, i.e. energy, is found from $p^{t}$ = $\sqrt{ m^2 - {{\overrightarrow{p}}}^2 }.$

The coefficient functions $u$ and $v$ are constrained by the ways quantities in (\ref{psi+}) transform under Poincar\'{e} transformations to a new spacetime reference frame: ({\it{i}}) the operators $a$ and $a^{\dagger}$ transform with a unitary representation (rep), ({\it{ii}}) the coefficients $u$ and $v$ are required to be invariant and ({\it{iii}}) the quantum field transforms by a nonunitary rep.

We part company with Weinberg by scattering the spacetime transformations of fields and states, using the equivalence of inertial frames to free the frame for fields from the frame for states. The reference frame for the fields undergoes a Poincar\'{e} transformation to a new frame, while the frame for the states and operators is transformed to their new frame.

Let the reference frame of the fields be related to the frame for the states and operators by a Lorentz transformation $\lambda.$ The new frame for the fields is obtained by applying a Lorentz transformation  $\Lambda,$ while the new state and operator frame is obtained with $\Lambda_{S},$ where $S$ indicates `States'. Call the coordinates $x$ for the field frame and $x_{S}$ for the state frame. We have
\begin{equation} \label{yTOx}
 x  = \lambda x_{S}  \quad {\mathrm{and}} \quad x^{\prime} = \Lambda x = \lambda x^{\prime}_{S}  \quad {\mathrm{and}} \quad	 \Lambda = \lambda \Lambda_{S} \lambda^{-1}	\, . %Eqn Check (E1)
\end{equation}
 The similarity transformation $\lambda$ relating $\Lambda_{S}$ and $\Lambda$ means that they are equivalent transformations, but in different inertial reference frames. 

That takes care of rotations and boosts, now for translations. When the fields are translated along the displacement $b$ to get to their new reference frame, the states are translated along some possibly different displacement $b_{S}.$ 

The description of an experiment depends on relative coordinates in which any global displacement, like $b$ or $b_{S},$ cancels out. Thus $b$ and $b_{S}$ can be completely arbitrary. Instead, assume a functional dependence so that the displacement $b_{S}$ for the states depends on the Lorentz transformation $\Lambda,$ the event $x,$ and the displacement of the field $b.$
\begin{equation} \label{bS}
 b_{S} = b_{S}(\Lambda,x,b)	\, . %Eqn Check (E2)
\end{equation}
Conventionally, the two displacements would agree, i.e. $b_{S} \rightarrow b.$

Thus, for a Poincar\'{e} transformation of the field $\psi,$ the operators $a$ and $a^{\dagger}$ (and states) transform by $( \Lambda_{S},b_{S})$ = $(  \lambda^{-1}\Lambda \lambda ,b_{S}(\Lambda,x,b)).$ The unitary transformation  $U( \Lambda_{S},b_{S})$ applied to operators yields 
\begin{equation} \label{Da+} %Eqn Check (D1,2)
 U(\Lambda_{S},b_{S}) \psi^{+}_{l}(x) {U}^{-1}(\Lambda_{S},b_{S}) =  \int d^3 p \enspace u_{l\sigma}(x,{\overrightarrow{p}})e^{i \Lambda_{S} p \cdot b_{S}} \sqrt{\frac{(\Lambda_{S} p)^t}{p^t}}   D^{(j)}_{\sigma \bar{\sigma}}(W^{-1})  a_{\bar{\sigma}}({\overrightarrow{\Lambda_{S} p}}) \, ,
\end{equation}
where the dot indicates the scalar product $p \cdot x \equiv$ $\eta_{\alpha \beta} p^{\alpha} x^{\beta}$ with flat spacetime metric $\eta,$ $D^{(j)}$ is a spin $j$ unitary representation of rotations, $W(\Lambda_{S},\overrightarrow{p})$ is the Wigner rotation for $\Lambda_{S}:$ $k \rightarrow$ $p \rightarrow$ $\Lambda_{S}p \rightarrow$ $k$ and $k$ = $(0,0,0,m).$ The coefficients $u$ are invariant.

Contrary to the unitary transformation $U(\Lambda_{S},b_{S})$ of operators, the fields transform with a nonunitary rep $D(\Lambda,b),$ 
\begin{equation} \label{Dpsi}
U(\Lambda_{S},b_{S}) \psi^{+}_{l}(x) {U}^{-1}(\Lambda_{S},b_{S}) =  D^{-1}_{l \bar{l}}(\Lambda,b)  \psi^{+}_{\bar{l}}(\Lambda x + b) \, ,		%Eqn Check (C5)
\end{equation}
where $\Lambda$ = $\lambda \Lambda_{S} \lambda^{-1}$ by (\ref{yTOx}) with $\lambda$ the state frame to field frame transformation and  $\Lambda x + b$ is the location in the new field frame of the event $x.$   

The expression for $\psi^{-}$ differs from the expression for $\psi^{+}$ only by $v$ for $u,$ $-i$ for $i,$ and $D^{(j)\ast}$ for $D^{(j)}.$ It makes no sense here to write expressions for both; henceforth consider mainly $\psi^{+}.$ The discussion is similar for $\psi^{-},$ except for some special considerations with $D^{(j)\ast},$ see \cite{W} for details.

Not every Lorentz rep $D(\Lambda,0)$ can be matched with any Poincar\'{e} rep $D(\Lambda,b),$ which includes a translation along $b.$ There are some requirements.\cite{Liubarsky,Shurtleff} The nontrivial translations sought here require reducible Lorentz reps $D(\Lambda,0)$. In standard notation, the spin of the Lorentz rep $D(\Lambda,0)$ must be of the form $(A,B)\oplus(C,D)$ with spins $(A,B)$ linked to $(C,D).$ Spins $(A,B)$ and $(C,D)$ are linked when 
\begin{equation} \label{ABCD}
(C,D) = (A\pm 1/2, B \pm 1/2) \, .
\end{equation}
For example, the Dirac 4-spinor, spin $(1/2,0) \oplus (0,1/2),$ has linked spins.
%Check References. 

Following the conventional process, the confluence of unitary and nonunitary transformations of operators and fields in the sums (\ref{psi+}) yields expressions for the coefficient functions $u(x,{\overrightarrow{p}}).$ The standard process is to write the fields on the left and right in (\ref{Dpsi}) with the sums of operators in (\ref{psi+}), transform the dummy variables $\overrightarrow{p}$ to $\overrightarrow{\Lambda_{S}p},$ and equate integrands. One gets
\begin{equation} \label{DuDu}
e^{i (\Lambda_{S}p)\cdot b_{S}}\sqrt{\frac{p^{t}}{(\Lambda_{S}p)^{t}}} D_{l \bar{l}}(\Lambda,b) u_{\bar{l} \sigma} (x, \overrightarrow{p}) =  u_{l \bar{\sigma}}(\Lambda x + b, \overrightarrow{\Lambda_{S}p}) D^{(j)}_{ \bar{\sigma} \sigma} \left(W(\Lambda_{s},\overrightarrow{p})\right)   \, ,		%Eqn Check (D5)
\end{equation}
where the $D^{(j)}$ and $D$ have changed sides to avoid the inverses seen in (\ref{Da+}) and (\ref{Dpsi}). 

Both $u_{\bar{l} \sigma} (x, \overrightarrow{p})$ and $u_{l \bar{\sigma}}(\Lambda x + b, \overrightarrow{\Lambda_{S}p})$ can be written in terms of the coefficient functions at the origin $x$ = 0. In (\ref{DuDu}) substitute $\Lambda$ = 1 and $b$ = $-x,$ so that $\Lambda x + b$ = 0, $\Lambda_{S}$ = 1, and $W$ = 1. We get
\begin{equation} \label{DuDuCase1}
 u_{\bar{l} \sigma} (x, \overrightarrow{p}) = e^{-i p\cdot b_{S}(1,x,-x)} D_{\bar{l} l }(1,+x) u_{l \sigma}(0, \overrightarrow{p})   \, .		%Eqn Check (E3)
\end{equation}
Then put $\Lambda$ = 1, with $x$ = $\tilde{\Lambda} \tilde{x} + \tilde{b}$ and $b$ = $-\tilde{\Lambda} \tilde{x} - \tilde{b},$ so that again we have $\Lambda x + b$ = 0, $\Lambda_{S}$ = 1, and $W$ = 1. Drop the tildes. This gives
$$
 u_{\bar{l} \sigma} (\Lambda x + b, \overrightarrow{p}) = e^{-i p\cdot b_{S}(1,\Lambda x + b,-\Lambda x - b)} D_{\bar{l} l }(1,\Lambda x + b) u_{l \sigma}(0, \overrightarrow{p})   \, ,		%Eqn Check (E5)
$$
or, by resetting the momentum $p,$ $p \rightarrow$ $\Lambda_{S}p,$
\begin{equation} \label{DuDuCase2}
 u_{\bar{l} \sigma} (\Lambda x + b, \overrightarrow{\Lambda_{S}p}) = e^{-i \Lambda_{S}p \cdot b_{S}(1,\Lambda x + b,-\Lambda x - b)} D_{\bar{l} l }(1,\Lambda x + b) u_{l \sigma}(0, \overrightarrow{\Lambda_{S}p})   \, .		%Eqn Check (E5)
\end{equation}
 Substituting expressions (\ref{DuDuCase1}) and (\ref{DuDuCase2}) back in (\ref{DuDu}) and taking steps to put all $x$- and $b$-dependence on the left side gives 
 $$
    e^{+i \Lambda_{S} p \cdot \left[b_{S}(\Lambda,x,b)-\Lambda_{S} b_{S}(1,x,-x) +  b_{S}(1, \Lambda x + b,-\Lambda x - b)\right] } D_{l \bar{l}}(\Lambda,0) u_{\bar{l}\sigma}(0,{\overrightarrow{p}}) = \hspace{5cm}
$$ 
\begin{equation} \label{Du4}\hspace{5cm} \sqrt{\frac{(\Lambda_{S}p)^{t}}{p^t}}  u_{l\bar{\sigma}}(0,{\overrightarrow{\Lambda_{S}p}}) D^{(j)}_{\bar{\sigma} \sigma}(W(\Lambda_{S},\overrightarrow{p}))    \, . %Eqn Check (F1,2,3)
\end{equation}
The exponent on the left shows the derivation history with $\Lambda$ = 1 and $b$ = $-x$ in $b_{S}(1,x,-x)$ and $\Lambda$ = 1 with $b$ = $-x$ = $-\tilde{\Lambda} \tilde{x} - \tilde{b}$ in $b_{S}(1, \Lambda x + b,-\Lambda x - b).$ 

Since $x$ and $b$ are confined to the left in (\ref{Du4}), the function $b_{S}(\Lambda,x,b)$ makes the  4-vector
$$ 
V(\Lambda,x,b) \equiv b_{S}(\Lambda,x,b)-\Lambda_{S} b_{S}(1,x,-x) +  b_{S}(1, \Lambda x + b,-\Lambda x - b) %Definition of V
$$
independent of $x$ and $b.$ To do this, one can show that $b_{S}(\Lambda,x,b)$ must be in the following form,
\begin{equation} \label{epsilon1}
 b_{S}^{\mu}(\Lambda,x,b) =  b_{0}^{\mu}(\Lambda) - \Lambda^{\mu}_{S \,\sigma} [A(x)]^{\sigma}_{\nu} x^{\nu} + [A(\Lambda x+b)]^{\mu}_{\nu} (\Lambda x+b)^{\nu} \, ,	%Eqn Check (F4)
\end{equation}
where $A(x)$ is an arbitrary second rank tensor field. By (\ref{epsilon1}), we have
$$ 
V(\Lambda,x,b) \equiv b_{0}(\Lambda)-\Lambda_{S} b_{0}(1)+ b_{0}(1) \, ,
$$
which is independent of $x$ and $b,$ as was required. For simplicity, drop the constant, $b_{0}^{\mu}$ = 0. %Eqn Check (F8)

 One recovers $b_{S}$ = $b$ when $\lambda$ = 1 and $A$ is the identity, $A^{\mu}_{\nu}$ = $\delta^{\mu}_{\nu},$ where $\delta$ is one for equal indices and zero otherwise. Also, note that $b_{S}$ depends on the values of the field $A$ at both $x$ and $\Lambda x+ b,$ the old and the new coordinates of the event. %Eqn Check (H1)

By the expression for $b_{S}^{\mu}(\Lambda,x,b)$ in (\ref{epsilon1}), we have $b_{S}^{\alpha}(1,x,-x)$ = $-A^{\alpha}_{\mu}x^{\mu}.$ Then $u_{\bar{l} \sigma} (x, \overrightarrow{p})$ in  (\ref{DuDuCase1}) becomes
\begin{equation} \label{Du8}
   u_{l\sigma}(x ,{\overrightarrow{p}}) = \sqrt{\frac{ m}{p^{\, t}}} \, e^{i p \cdot Ax}  D_{l \bar{l}}(L,x) u_{\bar{l}\sigma}(0,{\overrightarrow{0}})     \, .		%Eqn Check (G5)
\end{equation}
To get to the rest momentum $\overrightarrow{p}$ = $\overrightarrow{0},$ which is the 4-vector $k$ = $(0,0,0,m),$ we considered (\ref{Du4}) with $p$ = $k$ and with $\Lambda_{S}$ = $L_{S}$ = $\lambda^{-1} L \lambda$ is a transformation taking $k$ to $p$ so that $W(L_{S},\overrightarrow{0})$ = 1. 
%$b_{S}^{\alpha}(1,x,-x)$ = $-A^{\alpha}_{\mu}x^{\mu}.$: Eqn Check (G4)

 Substituting (\ref{Du8}) in the expression for $\psi^{+}_{l}(x),$  (\ref{psi+}) and the similar expression for $v_{l\sigma}(x,{\overrightarrow{p}})$ in for $\psi^{-}_{l}(x),$  determines many aspects of the quantum field $\psi_{l}(x).$ The complications of further developments along these lines\cite{W} are not followed here. Here we investigate the effects of the arbitrary field $A$ on the phase.

\section{Paths of Extreme Phase} \label{PhaseX}

Paths with extreme phase are the most likely, since deviations from the path introduces interference.\cite{Feynman} The most likely paths are expected to be the paths of particles in classical mechanics. Successfully describing classical motion is a first step in quantum theory.

Consider intervals $\delta x$ so short that $A$ varies negligibly along $\delta x.$ Also assume that $p$ is effectively constant along $\delta x.$ Then the change in the phase $\Theta$ associated with a particle of momentum $p$ is, by (\ref{Du8}), 
\begin{equation} \label{dTheta}
\delta \Theta = p \cdot A \delta x = \eta_{\alpha \beta}p^{\alpha} A^{\beta}_{\mu} \delta x^{\mu} \,  %Eqn Check (I1)
\end{equation}
 over such a displacement $\delta x.$

Let $\hat{p}$ be the timelike unit vector $ p/m$ and let the spacelike unit vector $\hat{p}_{\perp}$  have a null scalar product with $\hat{p}.$ Denote the magnitude of $A\delta x$ by $\delta \tau.$ In detail, $\eta_{\alpha \beta} \hat{p}^{\alpha} \hat{p}^{\beta}$ = $-1,$ $\eta_{\alpha \beta} \hat{p}^{\alpha}_{\perp} \hat{p}^{\beta}_{\perp}$ = $+1,$  $\eta_{\alpha \beta} (A\delta x)^{\alpha} (A\delta x)^{\beta}$ = $-\delta \tau^{2}.$ Thus, $A\delta x$ = $\delta \tau (\cosh{\phi} \hat{p}  + \sinh{\phi} \hat{p}_{\perp}),$ and $\delta \Theta$ = $-m \delta \tau \cosh{\phi}.$ The extreme value of $\cosh{\phi}$ occurs for $\phi$ = 0, so the extreme value of $\delta \Theta$ is given by $\delta \Theta_{\mathrm{extreme}}$ = $-m \delta \tau,$ and one has
\begin{equation} \label{pAdx2} A^{\alpha}_{\mu} {\delta X}^{\mu} = A^{\alpha}_{\mu} {\delta x}^{\mu}_{{\mathrm{extreme}}} =   m^{-1} p^{\alpha} \delta \tau   \, , \quad ({\mathrm{extreme}} \; \delta \Theta ) \quad  .  %Eqn Check (I10). Also the check for version 2 eqns (nI1-8) 7/1/15
 \end{equation}
The upper case ${\delta X}$ = ${\delta x}_{{\mathrm{extreme}}}$ indicates a path of extreme phase $\Theta.$ The extreme path occurs when  $A^{\alpha}_{\mu} {\delta X}^{\mu}$ is parallel to the momentum $p.$

To find a finite path of extreme phase $X(\tau)$ continue by attaching a second $\delta X,$ perhaps for slightly different $A$ and $p,$ to the $\delta X$ just found. The process forms a curve $X(\tau)$ of extreme phase with $A^{\alpha}_{\mu}(\tau)$ and $p^{\alpha}(\tau)$ the values of $A$ and $p$ along the curve $X(\tau).$

Indicating the derivative with respect to $\tau$ with a dot, 
\begin{equation} \label{Xdot}
\dot{X} \equiv  dX/d \tau \, ,
\end{equation}  %Eqn Check (I11)
(\ref{pAdx2}) becomes  
\begin{equation} \label{Ydot0}   p^{\alpha} = m A^{\alpha}_{\mu} \dot{ X}^{\mu}\, .   \quad ({\mathrm{Extreme}} \; {\mathrm{phase}}) \quad %Eqn Check (I10)
 \end{equation}
This equation relates the particle momentum $p$ to the velocity, the tangent $\dot{X},$ along the path for extreme phase. 

We can get `local coordinates' $\xi(x)$ if $A$ is the transformation field
\begin{equation} \label{Local}
A^{\alpha}_{\mu} = \frac{\partial \xi^{\alpha} }{ \partial x^{\mu}} \, . \quad \quad ({\mathrm{Local}} \; {\mathrm{coordinate}} \; {\mathrm{transformation}}) \quad %Eqn Check (H2)
\end{equation}
A neighborhood of  the event $x_{0},$ $x$ = $x_{0}+\delta x,$ is mapped into a neighborhood of $\xi(x_{0})$ by
$ \xi^{\alpha}( x)$ = $\xi^{\alpha}( x_{0}+\delta x)$ = $\xi^{\alpha}_{0} + \frac{\partial \xi^{\alpha}}{\partial x^{\mu}} \delta x^{\mu}$  = $\xi^{\alpha}_{0} +A^{\alpha}_{\mu} \delta x^{\mu},$		%Eqn Check (H4,5)
where higher order terms in $\delta x$ are dropped.

Not all tensor fields allow such an interpretation. The second order partials must commute, $\partial^{2} \xi^{\alpha}/\partial{x^{\lambda}}\partial{x^{\mu}}$ = $\partial^{2} \xi^{\alpha}/\partial{x^{\mu}}\partial{x^{\lambda}},$ or
\begin{equation} \label{symmDER}  \frac{\partial{A^{\alpha}_{\mu}}}{\partial{x^{\lambda}}} = \frac{\partial{A^{\alpha}_{\lambda}}}{\partial{x^{\mu}}} \, . \quad \quad  ({\mathrm{Integrability}} \; {\mathrm{conditions}}) \quad  %Eqn Check (H3)
\end{equation}
These are integrability conditions when, as is assumed, the field $A$ is a field of transformations.

By (\ref{Ydot0}) and (\ref{Local}), the curve $X(\tau)$ transforms to the curve $\Xi^{\alpha}(\tau)$ = $\xi^{\alpha}( X(\tau))$ of extreme phase in coordinates $\xi$ with 
\begin{equation} \label{pXi}
p^{\alpha} = m \dot{\Xi}^{\alpha} = m A^{\alpha}_{\mu} \dot{ X}^{\mu} \, ,
\end{equation}  
where uppercase $\Xi(\tau)$ denotes the path of extreme phase in local coordinates $\xi.$ %Eqn Check (I12) 

\section{Parallel Translations} \label{PT}

Next the concept of parallel translation of the momentum is introduced to determine how the momentum changes with location. Changing momentum with location alters the path $X(\tau)$ of extreme phase.

Momentum is a 4-vector. Rotations and boosts mix the components of the 4-vector by presumably familiar rotation and boost matrices, a spin $(1/2,1/2)$ representation of the Lorentz group. See, for example Ref. \cite{Tung} Chap. 10.

It is less well known, but true nevertheless, that a translation can change the momentum much as a rotation or boost.\cite{Liubarsky,Shurtleff} But, to do so, the momentum must be linked to another object with a suitable spin. The rule, presented previously in (\ref{ABCD}), requires an object with spin $(C,D)$ to be linked to vector spin $(A,B)$ = $(1/2,1/2)$ if $C$ = $A \pm 1/2$ and $D$ = $B \pm 1/2.$ Thus $(C,D) \in$ $\{(0,0),(1,0),(0,1),(1,1)\},$ which is the spin composition of a second rank tensor. Thus translation links the momentum to some second rank tensor $T$ together making a $4+16$ = 20-component quantity $\Phi,$
\begin{equation} \label{pT1}
 \Phi  = \pmatrix{p^{\alpha} \cr T^{\gamma \delta}_{(\Phi)} } \, .	%Eqn Check (J1)
\end{equation}
Like $A$, $T_{(\Phi)}$ is free, a collection of 16 free parameters, to be constrained as needed.

To generate the rotations and boosts of $\Phi,$ select a standard rep of the Lorentz group with angular momentum and boost generators. The choice here is from Ref. \cite{WeinbergV},
\begin{equation} \label{J11} 
 J^{\rho \sigma}  = -i \pmatrix{ \eta^{\sigma \mu} \delta^{\rho}_{\nu} - \eta^{\rho \mu} \delta^{\sigma}_{\nu}  && 0 \cr 0 && + \eta^{\rho \gamma} \delta^{\sigma}_{\epsilon} \delta^{\delta}_{\xi}- \eta^{\sigma \gamma} \delta^{\rho}_{\epsilon} \delta^{\delta}_{\xi} + \eta^{\rho \delta} \delta^{\sigma}_{\xi} \delta^{\gamma}_{\epsilon}- \eta^{\sigma \delta} \delta^{\rho}_{\xi} \delta^{\gamma}_{\epsilon}  }\, .   %Eqn Check (J2,3)
\end{equation} 
Then translations are generated with momentum matrices $P^{\mu},$\cite{RS}
\begin{equation} \label{Pmu}
 P^{\sigma} = \pmatrix{0 && P^{\sigma}_{12} \cr 0 && 0}  = i \pmatrix{ 0 && \pi_{1} \delta^{\sigma}_{\gamma} \delta^{\alpha}_{\delta}  + 
 \pi_{2}\delta^{\sigma}_{\delta} \delta^{\alpha}_{\gamma}  +  \pi_{3} \eta^{\sigma \alpha} \eta_{\gamma \delta}
  + \pi_{4}  \eta^{\sigma \rho} \eta^{\alpha \kappa} \epsilon_{\rho \kappa \gamma \delta} \cr 0 && 0  }  \, ,  %Eqn Check (J4)
\end{equation} 
where $\epsilon$ is the antisymmetric symbol and there are four constants $\pi_{i}$ because the transformation of a second rank tensor combines the four irreducible reps $\{(0,0),(1,0),(0,1),(1,1)\}.$  

There is another set of matrices that change $T_{(\Phi)}$ and leave $p$ unchanged. But we don't want that. The momentum matrices, displayed above with only the 12-block nonzero, change the four-vector momentum $p^{\alpha}$ and leave the tensor $T_{(\Phi)}$ unchanged. Keeping both blocks, so both $p^{\alpha}$ and $T_{(\Phi)}$ change,  makes the momentum matrices no longer commute, $[P^{\mu},P^{\nu}] \neq$ 0, and translations would not commute, which is deemed unacceptable spacetime behavior and violates the Poincar\'{e} algebra.\cite{W,Tung} The needed momentum matrices are those in (\ref{Pmu}).

Since the momentum matrices $P^{\mu}$ have nonzero components only in an off-diagonal block, the product of any two vanishes, $P^{\mu}P^{\nu}$ = 0, and any function of the $P^{\mu}$s that can be expanded in a power series reduces to a linear function. Thus the translation matrix for a translation along a displacement $\delta x$ is 
$$ Q(\delta x)=\exp{(- i  \delta x_{\sigma} P^{\sigma})} = {\mathbf{1}} - i \delta x_{\sigma} P^{\sigma} \, ,   %Eqn Check (J6)
$$
where ${\mathbf{1}}$ is the $20 \times 20$ unit matrix. The translation of a four-vector $v^{\, \alpha}$ yields $v^{\prime \, \alpha},$ with
\begin{equation} \label{Dp2}
{v^{\prime}}^{\alpha}  = \left[Q(\delta x)\right]^{\alpha}_{\sigma}v^{\sigma} = v^{\alpha}  - i \eta_{\sigma \mu}  (P^{\sigma}_{12})^{\alpha}_{\beta \gamma} T^{\beta \gamma}_{(\Phi)}  \delta x^{\mu} =  v^{\alpha}+ \eta_{\sigma \mu }  {T}^{\alpha \sigma}  \delta x^{\mu}  \, ,  %Eqn Check (J7)
\end{equation}
where the tensor $T$ is an abbreviation,
\begin{equation} \label{Tdef}
{T}^{\alpha \sigma} \equiv -i (P^{\sigma}_{12})^{\alpha}_{\beta \gamma} T^{\beta \gamma}_{(\Phi)}
  \, .   %Eqn Check (J8)
\end{equation}
Translation adds an inhomogeneous term $ T \delta x$ to $v.$ The added term is the same for any 4-vector $v$ even if the components of $v$ vanish. Also, translations are path dependent. A large displacement is the result of various sequences of small displacements, so over finite intervals the translated four-vector $v^{\prime}$ may depend on path. 

The question arises: What four-vector at $x+\delta x$ is equivalent to the four-vector at $x$? Is it the four-vector with the same components or the translated four-vector? We assume it is the translated four-vector, so that simple translation does not produce any innate change to the four-vector.

{\it{Parallel Translation of a Four-Vector.  The translated four-vector $v^{\prime}$ is {\it{equivalent}} to the original four-vector $v.$}}

The particle momentum is a 4-vector. The momentum at a nearby event that is equivalent to the momentum at an original event should be obtained by parallel translation.

{\it{Dynamical Postulate. A particle in a given eigenstate of momentum $p$ remains in eigenstates of equivalent momenta as spacetime is translated.}}

The coefficient functions $u_{l\sigma}(x ,{\overrightarrow{p}})$ in (\ref{Du8}) are referenced to the coefficient function at an `origin' $x$ = 0. Suppose the origin is on the semiclassical path of extreme phase $ X(\tau).$ As the particle moves along $X(\tau),$ its momenta change to  equivalent momenta. The interval $\delta X$ in (\ref{pAdx2}) for the extreme phase change $\delta \Theta = p \cdot A \delta X$ is followed by the interval $\delta X^{\prime}$ obtained for the extreme phase change $\delta \Theta^{\prime} = p^{\prime} \cdot A \delta X^{\prime}$ with the momentum $p^{\prime}$ obtained by parallel translation along $\delta X,$ a special case of (\ref{Dp2}). 

Therefore, the translated momentum $p^{\prime}$ is 
\begin{equation} \label{Dp}
{p^{\prime}}^{\alpha}  =  p^{\alpha} + \eta_{\sigma \mu }  {T}^{\alpha \sigma}  \delta X^{\mu}  \, .  %Eqn Check (J7)
\end{equation}  
Since the path $X(\tau)$ is a succession of intervals and the momentum is parallel translated along each interval, the momentum is a function $p(\tau)$ of proper time $\tau,$ whose derivative with respect to $\tau$ is determined by parallel translation (\ref{Dp}) to be 
 \begin{equation} \label{Dp3a}
\dot{ p}^{\alpha} =  \eta_{\sigma \mu} {T}^{\alpha \sigma } {\dot{X}}^{\mu}  \, .  \quad ({\mathrm{Parallel}} \; {\mathrm{translation}}) \quad  %Eqn Check (J10)
\end{equation}
The tensor $T$ is a free parameter that is constrained by the other equations satisfied by the momentum $p$ and the path of extreme phase $X.$
Equation (\ref{Dp3a}) is the semi-classical equation of motion.

%\pagebreak

\section{Curved Spacetime} \label{curvedST}

Curved spacetime is a consequence of extreme phase and the fact that mass is the magnitude of the momentum.  Substitute the requirement of extreme phase $p$ = $mA\dot{X},$ by (\ref{Ydot0}), into the mass equation,  
\begin{equation} \label{p2m2}
\eta_{\alpha \beta} p^{\alpha}p^{\beta} = -m^{2} \, , %Eqn Check (A) 
\end{equation} 
$$  
\frac{1}{m^2}\eta_{\alpha \beta} p^{\alpha}p^{\beta}  = \eta_{\alpha \beta} A^{\alpha}_{\mu}A^{\beta}_{\nu}\dot{ X}^{\mu}\dot{ X}^{\nu} =   -1    \, .   
$$
By collecting some of the quantities together in $g_{\mu\nu}$, one has
\begin{equation} \label{g2}  g_{\mu \nu} \dot{ X}^{\mu}\dot{ X}^{\nu} =   -1    \, ,  %Eqn Check (K1)
\end{equation}
where the tensor field $g_{\mu \nu}(x)$ and its inverse are defined in terms of the tensor field $A(x)$ by
\begin{equation} \label{g3}  g_{\mu \nu} \equiv  \eta_{\alpha \beta} A^{\alpha}_{\mu}A^{\beta}_{\nu} \quad {\mathrm{and}} \quad g^{\mu \nu} \equiv  \eta^{\alpha \beta} {A}_{\; \alpha}^{-1 \; \mu}A_{\; \beta}^{-1 \; \nu}\, ,  %Eqn Check (K2,3)
\end{equation}
with $g_{\mu \sigma}g^{\sigma \rho}$ = $\delta^{\rho}_{\mu},$ as is easily verified.  Since the flat spacetime metric $\eta_{\alpha \beta}$ is symmetric, both $g_{\mu \nu}$ and $g^{\rho \sigma}$ are symmetric. We call $g_{\mu \nu}$ the `curved spacetime metric'. 

By (\ref{g2}), (\ref{g3}), and  $\dot{\Xi}^{\alpha}$ = $ A^{\alpha}_{\mu} \dot{ X}^{\mu}$ from (\ref{pXi}), one sees that
\begin{equation} \label{xi2}  \eta_{\alpha \beta} \dot{\Xi}^{\alpha} \dot{\Xi}^{\beta} = -1 \, ,  %Eqn Check (K7)
\end{equation}
where $\Xi(\tau)$ is the path of extreme phase in coordinates $\xi.$ 

Because the metric in (\ref{xi2}) is the flat spacetime metric $\eta,$ the transformation $A$ gives locally flat spacetime coordinates $\xi$ from curved spacetime coordinates $x.$ It follows that $g_{\mu \nu}$ is locally Lorentzian.

Having two metrics makes raising and lowering indices a possible source of confusion. Thus raising and lowering indices is kept to a minimum and, when needed, the metric involved is displayed clearly.

Furthermore, we can transform the path $\Xi(\tau)$ at the event {\sc{O}} at $\tau$ = 0 to a frame so that $\dot{\Xi}(0)$ has only its time component nonzero, $\dot{\Xi}(0)$ = $(0,0,0,\dot{\Xi}_{0}^{t}).$ Then, by (\ref{xi2}), we have $\dot{\Xi}_{0}^{t}$ = 1 and $d\Xi^{t}_{0}$ = $d\tau,$ so  we are again justified in calling the quantity $\tau$ {\it{the `proper time' along the path $X(\tau)$ of extreme phase.}} The parameter $\tau$ is the time in a local Lorentz frame in which the particle is momentarily at rest.

Turn now to the particle momentum. Define a 4-vector $\bar{p}^{\mu}(\tau)$ by 
\begin{equation} \label{pbar}
 \bar{p}^{\mu} \equiv m \dot{X}^{\mu} \, .    %Eqn Check (K8)
\end{equation}
By (\ref{g2}), one finds that 
\begin{equation} \label{gm2a}  g_{\mu \nu} {\bar{p}}^{\mu}{\bar{p}}^{\nu} = -m^{2} \, .  %Eqn Check (K10)
\end{equation}
Thus the mass $m$ is the magnitude of the momentum ${\bar{p}}$ calculated with the metric $g$ and we see that $\bar{p}^{\mu}(\tau)$ is the `curved spacetime momentum' of the particle along the path $X(\tau)$ of extreme phase. And $p^{\alpha}$ is the flat spacetime particle momentum because of the flat spacetime metric $\eta$ in (\ref{p2m2}), $\eta_{\alpha 
\beta}p^{\alpha} p^{\beta}$ = $-m^2.$ By (\ref{Ydot0}) and (\ref{pbar}), it follows that 
\begin{equation} \label{gm2}
p^{\alpha} = A^{\alpha}_{\mu} \bar{p}^{\mu} \, . %Eqn Check (K9)
\end{equation}
Thus the flat and curved spacetime momenta are related by the same transformation $A$ that, by (\ref{pXi}), relates the tangents of the flat spacetime and curved spacetime paths $\dot{\Xi}$ and $\dot{X}.$ 

%\pagebreak
\section{Combined Electromagnetic/Gravitational Motion} \label{EAG}

In this section, the semi-classical equation of motion (\ref{Dp3a}) is shown to describe the motion of a charged particle in gravitational and electromagnetic fields.

The path of extreme phase $X^{\mu}$ is the `curved' spacetime path because of the curved spacetime metric $g$ in $g_{\mu \nu} \dot{X}^{\mu} \dot{X}^{\nu}$ = $-1,$ (\ref{g2}). And the momentum $p^{\alpha}$ is the flat spacetime particle momentum because of the flat spacetime metric $\eta$ in the mass equation (\ref{p2m2}),
$\eta_{\alpha \beta} p^{\alpha}p^{\beta}$ = $-m^{2}.$ Thus the equation of motion (\ref{Dp3a}), $\dot{ p}^{\alpha}$ =  $\eta_{\sigma \mu} {T}^{\alpha \sigma } {\dot{X}}^{\mu},$ mixes the flat spacetime particle momentum $p^{\alpha}$ and the curved spacetime path $X^{\mu}.$ 

In (\ref{gm2}),   $p^{\alpha}$ is expressed in terms of the curved spacetime momentum, $p^{\alpha}$ = $A^{\alpha}_{\mu}\bar{p}^{\mu},$ and by substituting in the equation of motion (\ref{Dp3a}), we can arrange to have an equation of motion with the curved spacetime quantities ${\bar{p}}^{\mu}$ and $X^{\mu}.$ One finds that
 \begin{equation} \label{Dp3b}
\ddot{X}^{\mu} = \frac{1}{m}{\dot{{\bar{p}}}}^{\;\mu} =   \frac{1}{m}  A^{-1 \; \mu}_{\alpha} \eta_{\sigma \nu} {T}^{\alpha \sigma } {\dot{X}}^{\nu} -  A^{-1 \; \mu}_{\beta} \frac{\partial A^{\beta}_{\nu}}{\partial x^{\lambda}} \dot{X}^{\lambda}  {\dot{X}}^{\nu}  \, ,  %Eqn Check (K11) and 7/1/15 Check (nk3)
\end{equation}
where we used (\ref{Ydot0}), $p^{\alpha}$ = $m A^{\alpha}_{\mu} \dot{ X}^{\mu},$ to write $p$ in terms of $\dot{X}.$ 

The term quadratic in velocity, $ \dot{X}^{\lambda}  {\dot{X}}^{\mu}, $ looks a bit like the acceleration in a gravitational field. Write
\begin{equation} \label{Xddot2}  
  A^{-1 \; \mu}_{\beta} \frac{\partial A^{\beta}_{\nu}}{\partial x^{\lambda}} \dot{X}^{\lambda}  {\dot{X}}^{\nu} =    \frac{1}{2}  A^{-1 \; \nu}_{\beta} \left(\frac{\partial A^{\beta}_{\nu}}{\partial x^{\lambda}} + \frac{\partial A^{\beta}_{\lambda}}{\partial x^{\nu}}\right) \dot{X}^{\lambda}  {\dot{X}}^{\nu} =  C^{\mu}_{\lambda \nu}  \dot{X}^{\lambda}  {\dot{X}}^{\mu} \, ,   %Eqn Check (M2,M3)
\end{equation} where $C$ is given by
\begin{equation} \label{Cdef}
C^{\mu}_{\lambda \nu} =  \frac{1}{2}  A^{-1 \; \mu}_{\beta} \left(\frac{\partial A^{\beta}_{\nu}}{\partial x^{\lambda}} + \frac{\partial A^{\beta}_{\lambda}}{\partial x^{\nu}}\right) \, .
\end{equation}
Clearly, the quantity $C^{\mu}_{\lambda \nu}$ is symmetric in its lower indices, $C^{\mu}_{\lambda \nu}$ = $C^{\mu}_{\nu \lambda}.$ 

We now show that $C$ is indeed the Christoffel connection of the metric $g.$ The Christoffel symbol of the first kind is defined to be \cite{Adler}
$$ \left[\mu \nu, \rho   \right]  \equiv \frac{1}{2} \left( \frac{\partial g_{\mu \rho}}{\partial x^{\nu}}  + \frac{\partial g_{\nu \rho}}{\partial x^{\mu}}  -  \frac{\partial g_{\mu \nu}}{\partial x^{\rho}} \right) \, .  %Eqn Check (M3a)
$$
By the definitions of $C^{\mu}_{\lambda \nu}$ in (\ref{Cdef}) and $g_{\mu \nu}$ in (\ref{g3}), i.e. $g_{\mu \nu}$ = $\eta_{\alpha \beta} A^{\alpha}_{\mu}A^{\beta}_{\nu},$ one can show that
$$   \left[\mu \nu, \rho   \right]  = g_{\rho \sigma} C^{\sigma}_{\mu \nu} + \frac{1}{2} \eta_{\alpha \beta} \left[ A^{\alpha}_{\mu} \left( \frac{\partial{A^{\beta}_{\rho}}}{\partial{x^{\nu}}} - \frac{\partial{A^{\beta}_{\nu}}}{\partial{x^{\rho}}}\right) + A^{\alpha}_{\nu} \left( \frac{\partial{A^{\beta}_{\rho}}}{\partial{x^{\mu}}} - \frac{\partial{A^{\beta}_{\mu}}}{\partial{x^{\rho}}}\right)\right] \, .  %Eqn Check (M6)
$$
Looking at this, we see that $C$ is the Christoffel connection when the terms in parentheses vanish, i.e. when
$$ \frac{\partial{A^{\beta}_{\rho}}}{\partial{x^{\nu}}} = \frac{\partial{A^{\beta}_{\nu}}}{\partial{x^{\rho}}} \, .  %Eqn Check (M7)
$$ 
But this is just the integrability condition (\ref{symmDER})  so that 
$ A^{\beta}_{\nu}$ = $\partial{\xi}^{\beta}/\partial{x}^{\nu} $
with $A^{\beta}_{\nu}$ transforming coordinates $x$ to $\xi^{\beta}(x)$ as in (\ref{Local}). %Eqn Check (H2)

Thus, when $A$ is a field of transformations to local coordinates, the quantity $C$ is the Christoffel connection of the metric $g,$ \cite{Adler}
\begin{equation} \label{CisCristoffel}
 C^{\sigma}_{\mu \nu} =  g^{\rho \sigma}\left[\mu \nu, \rho   \right] = \frac{g^{\rho \sigma}}{2} \left( \frac{\partial g_{\mu \rho}}{\partial x^{\nu}}  + \frac{\partial g_{\nu \rho}}{\partial x^{\mu}}  -  \frac{\partial g_{\mu \nu}}{\partial x^{\rho}} \right) \, .  %Eqn Check (M8)
\end{equation}
The `covariant derivative' of $\dot{X}$ with respect to $\tau$ is defined to be\cite{WeinbergV,covDER}
\begin{equation} \label{D2x3}
\frac{D\dot{ X}^{\mu}}{d\tau} \equiv  \ddot{ X}^{\mu} +  C^{\mu}_{ \lambda \nu} \dot{X}^{\lambda} {\dot{X}}^{\nu}  \, .  %Ref: Weinberg p. 103 (4.6.4) and Dirac p. 19 (10.7)
\end{equation}
With this, the equation of motion (\ref{Dp3b}) with (\ref{Xddot2}) can be written  as
 \begin{equation} \label{D2x4}
\frac{D\dot{ X}^{\mu}}{d\tau} = \frac{1}{m}  A^{-1 \; \mu}_{\alpha} \eta_{\sigma \nu} {T}^{\alpha \sigma } {\dot{X}}^{\nu}   \, . %ovbious, so double check it.
\end{equation}
The covariant derivative on the left has the same form in any coordinate system.

If the arbitrary tensor field $T$ is chosen properly, one can have the term on the right in (\ref{D2x4}) be the electromagnetic Lorentz force divided by mass $m$. According to Ref. \cite{WLf}, we need $T$ to satisfy
\begin{equation} \label{Fdef}
A^{-1 \; \mu}_{\alpha} \eta_{\sigma \nu} {T}^{\alpha \sigma }{\dot{X}}^{\nu} = q g_{\sigma \nu} F^{\mu \sigma} {\dot{X}}^{\nu} \, %Eqn check for v2 (nL1) 6/27/15
\end{equation}
for a particle of charge $q$ in an electromagnetic field $F.$ Thus the tensor field $T$ in (\ref{Tdef}) should be
\begin{equation} \label{TforEMcurved}
\eta_{\sigma \nu}{T}^{\alpha \sigma } = q A^{\alpha}_{\lambda}g_{\rho \nu} F^{\lambda \rho} \, .
\end{equation}
The formula simplifies when the displacements $b$ and $b_{S}$ for fields and states are equal, which is the conventional assumption. By (\ref{epsilon1}) and (\ref{g3}), that happens when $A^{\alpha}_{\lambda}$ = $\delta^{\alpha}_{\lambda}$ and $g_{\rho \nu}$ = $\eta_{\rho \nu},$ so that $T$ simplifies to $T^{\alpha \sigma}$ = $q F^{\alpha \sigma}.$
%WLf:  Wgrav p123 eq(5.1.11) & p125 eq(5.2.9,10)

Now the equation of motion (\ref{Dp3b}) is
 \begin{equation} \label{D2x4a}
\frac{D\dot{ X}^{\mu}}{d\tau} = \frac{q}{m}   g_{\sigma \nu} F^{\mu \sigma} {\dot{X}}^{\nu}   \, , 
\end{equation}
which is the Lorentz force equation in curved spacetime.\cite{WLf} 
This equation is covariant; it has the same same form when the metric $g$ is transformed to some other metric $g^{\prime}.$\cite{Misner} 

Since (\ref{D2x4}) is covariant, it can be transformed from $x$-coordinates to local $\xi$-coordinates via the transformation field $A^{\alpha}_{\nu}$ = $\partial{\xi^{\alpha}}/\partial{x^{\nu}}.$ One finds that
\begin{equation} \label{mddXi}
\ddot{\Xi}^{\alpha} =  \frac{q}{m} \eta_{\beta \kappa} f^{\alpha \beta} {\dot{\Xi}}^{\kappa} \, , %Eqn check for v2 (nL2) 6/27/15
\end{equation}
where $\ddot{\Xi}^{\alpha}$ = $d \dot{\Xi}^{\alpha}/d\tau,$ $\dot{\Xi}^{\alpha}$ = $A^{\alpha}_{\nu}\dot{ X}^{\nu},$ $f^{\alpha \beta}$ = $A^{\alpha}_{\nu}A^{\beta}_{\sigma}F^{\nu \sigma},$ and we used $g_{\sigma \nu}$ = $\eta_{\alpha \beta} A^{\alpha}_{\nu}A^{\beta}_{\sigma}$ by (\ref{g3}).

While the fields $f$ and $F$ have been called the electromagnetic field in local and general coordinates, it has not yet been shown that they must be antisymmetric. This will be undertaken now.

Take the derivative of $\eta \dot{\Xi} \dot{\Xi}$ = $-1,$ (\ref{xi2}), with respect to proper time $\tau$, 
$$ 0 = 2 \eta_{\alpha \beta} \dot{\Xi}^{\alpha} \ddot{\Xi}^{\beta} = \frac{2}{m} \eta_{\alpha \beta} \dot{\Xi}^{\alpha} q \eta_{\sigma \mu} f^{\beta \sigma}  {\dot{\Xi}}^{\mu} \, ,
$$
so that, after multiplying by $m/q,$
$$ 0 = 2\left(\eta_{\alpha \beta} \dot{\Xi}^{\alpha}\right) \left(\eta_{\mu \sigma} {\dot{\Xi}}^{\mu} \right)f^{\beta \sigma} =  \dot{\Xi}_{\beta} \dot{\Xi}_{\sigma}\left(f^{\beta \sigma}+f^{\sigma \beta}\right) \, ,  %Eqn Check (N4)
$$
where $\dot{\Xi}_{\beta}$ = $\eta_{\alpha \beta} \dot{\Xi}^{\alpha}.$ Thus $f$ is antisymmetric along the path of extreme phase. 
Since $f^{\alpha \beta}$ = $A^{\alpha}_{\nu}A^{\beta}_{\sigma}F^{\nu \sigma},$ if $f$ is antisymmetric, then $F$ is antisymmetric as well,
\begin{equation} \label{fFanti}
f^{\beta \sigma} = - f^{\sigma \beta} \quad {\mathrm{and}} \quad F^{\mu \nu} = - F^{\nu \mu} \, .
\end{equation}
The antisymmetry of $f$ and $F$ adds evidence to their identification as the electromagnetic field in local and general coordinates. The remaining problem of relating electromagnetic and gravitational fields to sources may be treated elsewhere.

\pagebreak

\end{document}